\documentclass[useAMS,usenatbib]{mn2e}

\def\Msun {\,\mathrm{M}_\odot}

\defcitealias{kinugawa16}{K16}
\defcitealias{kinugawa14}{K14}
\defcitealias{belczynski16}{B16}
\defcitealias{woosley16}{Woosley~2016}
\defcitealias{mink15}{dMB15}

\usepackage{amsmath}
\usepackage{amssymb}
\usepackage{graphicx}
\usepackage{color}
\usepackage{url}
\title[GWs from the Remnants of the First Stars]{Gravitational waves from the remnants of the first stars}

\author[T.\ Hartwig et al.]
{\parbox{\textwidth}{
Tilman Hartwig$^{1}$\thanks{E-mail: hartwig@iap.fr}, Marta Volonteri$^{1}$, Volker Bromm$^{2}$, Ralf S.\ Klessen$^{3}$,\\
Enrico Barausse$^{1}$, Mattis Magg$^3$, and Athena Stacy$^4$\\
}\\
$^{1}$Sorbonne UniversitŽ\'es, UPMC et CNRS, UMR 7095, Institut d'Astrophysique de Paris, 98 bis bd Arago, F-75014 Paris, France\\
$^{2}$Department of Astronomy, University of Texas, Austin, Texas 78712, USA\\
$^{3}$Universit\"at Heidelberg, Zentrum f\"ur Astronomie, Institut f\"ur Theoretische Astrophysik, Albert-Ueberle-Str.\ 2,\\
D-69120 Heidelberg, Germany\\
$^4$Department of Astronomy, University of California, Berkeley, CA 94720, USA
}

\begin{document}


\pagerange{\pageref{firstpage}--\pageref{lastpage}} \pubyear{2016}

\maketitle

\label{firstpage}

\begin{abstract}
Gravitational waves (GWs) provide a revolutionary tool to investigate yet unobserved astrophysical objects. Especially the first stars, which are believed to be more massive than present-day stars, might be indirectly observable via the merger of their compact remnants. We develop a self-consistent, cosmologically representative, semi-analytical model to simulate the formation of the first stars.
By extrapolating binary stellar-evolution models at 10\,\% solar metallicity to metal-free stars, we track the individual systems until the coalescence of the compact remnants.
We estimate the contribution of primordial stars to the merger rate density and to the detection rate of  the Advanced Laser Interferometer Gravitational-Wave Observatory (aLIGO). Owing to their higher masses, the remnants of primordial stars produce strong GW signals, even if their contribution in number is relatively small. We find a probability of $\gtrsim1\%$ that the current detection GW150914 is of primordial origin. We estimate that aLIGO will detect roughly 1 primordial BH-BH merger per year for the final design sensitivity, although this rate depends sensitively on the primordial initial mass function (IMF). Turning this around, the detection of black hole mergers with a total binary mass of $\sim\,300\Msun$ would enable us to constrain the primordial IMF. 
\end{abstract}


\section{Introduction}
The first detection of gravitational waves (GWs) on 2015 September 14, has opened a completely new window to investigate astrophysical processes and phenomena, which are otherwise invisible to observations in the electromagnetic spectrum \citep[but see][]{loeb16,perna16}. This first event GW150914 was the inspiral and merger of two black holes (BHs) with masses $M_1=36^{+5}_{-4}\Msun$ and $M_2=29^{+4}_{-4}\Msun$ at redshift $z=0.09^{+0.03}_{-0.04}$ \citep{PhysRevLett.116.061102}. It was detected by the Advanced Laser Interferometer Gravitational-Wave Observatory (aLIGO) with a false alert probability of $<2 \times 10^{-7}$ \citep{PhysRevLett.116.061102}. The local merger rate density inferred from this event is $2-400\,\mathrm{yr}^{-1}\mathrm{Gpc}^{-3}$ for BH--BH mergers \citep{LIGOrates}. 
GW150914 also indicates that the stochastic GW background could be higher than previously expected and potentially measurable by the aLIGO/Virgo detectors operating at their final sensitivity \citep{schneider00,LIGObackground}.

The detection probability increases with the mass of the merging BHs. The first, so-called Population~III (Pop~III), stars are believed to be more massive than present-day stars and yield consequently more massive remnants \citep[for reviews see][]{bromm13,glover13,greif15}. Due to their high masses, a significant fraction of the possible detections might originate from these primordial stars \citep{belczynski04,kulczycki06,kinugawa14,kinugawa16} and \citet{dominik13,dominik15} show that most of the binary BHs that merge at low redshift, have actually formed in the early Universe. Hence, it is worth investigating the contribution from Pop~III stars in more detail with a self-consistent model of primordial star formation.

In this Letter, we apply a semi-analytical approach to determine the rate density and the detection rate of mergers for aLIGO that originate from the first stars.\footnote{Our catalogues of Pop~III binaries are publicly available here:
\url{http://www2.iap.fr/users/volonter/PopIIIGW}}



\section{Methodology}
\label{sec:methods}
\subsection{Self-consistent Pop~III star formation}
We create a cosmologically representative sample of dark matter merger trees with the {\sc galform} code based on \cite{parkinson08}. The merger trees start at $z_{\rm max}=50$ and follow Pop~III star formation down to $z=6$, after which we do not expect significant Pop~III star formation to occur. The formation of primordial stars is modelled self-consistently, taking into account radiative and chemical feedback. We briefly review the main aspects of the model; see \citet{hartwig15,hartwig16} for details.

To form Pop~III stars, a halo has to be metal free and have a virial temperature high enough for efficient cooling by H$_2$. Moreover, we check the time-scale of dynamical heating due to previous mergers and the photodissociation of H$_2$ by external Lyman--Werner radiation. Once a dark matter halo passes these four criteria, we assign individual Pop~III stars to it, based on random sampling of a logarithmically flat initial mass function (IMF) in the mass range between $M_\mathrm{min} = 3\Msun$ and $M_\mathrm{max}=300\Msun$, motivated by, e.g., \citet{greif11}, \citet{clark11}, and \citet{dopcke13}. The IMF is still uncertain, and therefore besides this fiducial model, we also consider a low mass ($1-100\Msun$) and a high mass ($10-1000\Msun$) IMF.  The total stellar mass per Pop~III-forming halo is set by the star formation efficiency. This parameter is calibrated to reproduce the optical depth to Thomson scattering of $\tau = 0.066 \pm 0.016$ \citep{planck15}, taking also into account the contribution by later generations of stars, based on the global cosmic star formation history \citep{behroozi15}.


\subsection{Binary sampling and evolution}
We use the results of the most detailed study of Pop~III binary systems to date by \citet{stacy13}. They performed a cosmological simulation initialized at $z=100$ within a 1.4~Mpc (comoving) box. This simulation followed the eventual formation and evolution of Pop III multiple systems within 10 different minihaloes at a resolution of 20~au.
From this study, we adopt a binary fraction of 36\%, which translates into a $\sim\,50$\% probability for a single star to have a binary companion \citep[see also][]{stacy16}. Note that other studies of Pop~III star formation might allow larger binary fractions \citep{clark11,greif11,smith11}, but all derived merger and detection rates scale linearly with this binary fraction. Our results can thus readily be rescaled accordingly.

The evolution of the binary system and consequently the nature (BH or NS), the masses, and the time of coalescence of the two compact objects depend mainly on the zero-age main-sequence (ZAMS) characteristics, respectively, the semimajor axis and eccentricity of their orbit and their masses.
For the pairing of the binaries and the underlying distribution of mass ratios, we apply the `ordered pairing' advocated by \citet{oh15}, since observations show that massive binaries favour members with similar masses. Hence, we order the primordial stars in one halo by descending mass, check probabilistically if they have a binary companion, and pair the most massive with the second most massive, the third most massive with the fourth most massive and so on.
For the ZAMS eccentricity $e_0$ of each binary system we draw a random value from the thermal distribution $p(e)\,\mathrm{d}e \propto e\,\mathrm{d}e$ with $e_\mathrm{min}=0.1$ and $e_\mathrm{max}=1$ \citep[][hereafter \citetalias{kinugawa14}]{kroupa95,dominik12,kinugawa14}. This distribution agrees qualitatively with that in \citet{stacy13}.
The ZAMS semimajor axis $a_0$ is sampled from the distribution $p(x)\,\mathrm{d}x \propto x^{-1/2}\,\mathrm{d}x$ with $x=\log (a_0/\mathrm{R}_\odot)$, $a_\mathrm{min}=50\,\mathrm{R}_\odot$, and $a_\mathrm{max}=2 \times 10^6\,\mathrm{R}_\odot$. The shape and the lower limit are motivated by \citet{sana12} and \citet{mink15} (hereafter \citetalias{mink15}), whereas the upper limit is chosen in agreement with the data by \citet{stacy13}. We have verified that the specific choice of these limits does not significantly affect the final results.

Once we have identified the binaries and assigned their ZAMS quantities, we use the tabulated models for stellar binary evolution by \citetalias{mink15} to calculate the masses of the remnants and their delay time $t_\mathrm{del}$ until coalescence. The delay time is the sum of the time from the ZAMS to the formation of the last compact object and the ensuing inspiral time, $t_\mathrm{insp}$.  We chose their model `N-m2 A.002' (with the lowest available metallicity of 10\% solar), as the best fit to the properties of Pop~III stars in terms of IMF, metallicity, and evolutionary channels (see also \citetalias{kinugawa14}; \citet{belczynski16}, hereafter \citetalias{belczynski16})

Notice the differences between the stellar binary evolution of metal-free and metal-enriched systems, which might lead to systematic errors \citepalias{kinugawa14,kinugawa16}. In contrast to metal-enriched stars, Pop~III stars lose only a small fraction of their mass due to stellar winds, which yields more massive remnants and closer binaries, since the binding energy is not carried away by the winds. Moreover, Pop~III stars with a stellar mass of less than $\sim\,50\Msun$ evolve as blue supergiants (not as red supergiants, like metal-enriched stars) and the resulting stable mass transfer makes the common envelope phase less likely.

The data is tabulated for stellar masses of the individual companions of up to $150\Msun$. For higher masses, we proceed in the following way. We ignore binaries with one star in the mass range $140\Msun \leq M_* \leq 260\Msun$, as we do not expect any compact remnants due to pair-instability supernova (PISN) explosions \citep{heger02}. For stars above $260\Msun$, we consider $t_\mathrm{del}$ characteristic of stars with $100\Msun \leq M_* \leq 140\Msun$, adopt the final BH masses of primordial stars from \citet{heger02}, and correct the tabulated inspiral time according to \citepalias{kinugawa14}
$t_\mathrm{insp} \propto {m_1}^{-1} {m_2}^{-1} (m_1+m_2)^{-1}$,
where $m_1$ and $m_2$ are the masses of the binary compact objects. This approach is justified because the tabulated $t_\mathrm{del}$ show negligible dependence on stellar mass for massive stars.
 

\subsection{Detectability}
Based on the cosmologically representative, self-consistent sampling of Pop~III stars and the corresponding $t_\mathrm{del}$ of each binary, we determine the intrinsic merger rate density $R$. This represents the number of compact binary mergers per unit source time and per comoving volume, and is also referred to as the rest-frame merger rate density. To estimate the aLIGO detection rate, we calculate the single-detector signal-to-noise ratio (SNR) $\rho$ for each merger via~\citep{Maggiore:1900zz,1993PhRvD..47.2198F,1994PhRvD..49.2658C},
\begin{equation}
\rho^2=4 \int_0^{\infty} \frac{|h(f)|^2}{S_n(f)} \mathrm{d}f\,,
\end{equation}
where $h(f)$ is the Fourier-domain (sky- and orientation-averaged) 
GW strain at the detector, and $S_n$ is the noise power spectral density of a single aLIGO detector.
We assume that an event is detectable if $\rho>8$, as conventionally done in the LIGO literature \citep{ligo10,dominik15,2015arXiv151004615B,mink16}. (This translates to SNR larger than 12 for a three-detector network, e.g. the two aLIGOs and advanced Virgo.) 
For the current aLIGO detectors, we use the O1  noise power spectral density \citep{O1}, whereas
to assess detectability when the detectors are in their final design configurations we use the  zero-detuning, high-power
configuration of \cite{aligo_noise}. For $h(f)$, we use either inspiral-only, restricted post-Newtonian 
waveforms \citep[computing the Fourier transform with the stationary phase approximation, see e.g.][]{Maggiore:1900zz}, or inspiral-merger-ringdown PhenomA (non-spinning) waveforms~\citep{2008PhRvD..77j4017A,PhysRevD.79.129901}. We employ the former for BH--NS, and NS--NS systems, with
a cut-off at the frequency of the innermost stable circular orbit (ISCO; note that the ISCO frequency also
corresponds approximately to the merger frequency of NS--NS systems). For BH--BH systems, particularly at high masses,  the merger-ringdown contains considerable SNR, hence we use PhenomA waveforms. 
We then calculate the detection rate as~\citep{haehnelt94}
\begin{equation}
\frac{dn}{dt} = 4 \pi c {\int}_{\begin{subarray}{l}\rho >8\\
   z<z_{\max}\end{subarray}}
\mathrm{d}z \mathrm{d}m_1 \mathrm{d}m_2 \frac{\mathrm{d}^2R}{\mathrm{d}m_1 \mathrm{d}m_2}
\frac{\mathrm{d}t}{\mathrm{d}z}\left( \frac{d_{L}}{1+z} \right)^2,
\end{equation}
where the luminosity distance $d_{L}$ and the derivative of the look-back time with respect to $z$, $\mathrm{d}t/\mathrm{d}z$, are computed with a $\Lambda$CDM cosmology, and the integral is restricted to detectable events only ($\rho>8$). 

Finally, we characterize the stochastic GW background of our binary population by the energy density spectrum \citep[see e.g.][]{2001astro.ph..8028P,2011PhRvD..84h4004R,LIGObackground}:
\begin{equation}
\Omega_{\rm GW}(f)= \frac{f}{\rho_c c^2}\int_{z<z_{\max}} \mathrm{d}z \mathrm{d}m_1 \mathrm{d}m_2 \frac{\mathrm{d}^2R}{\mathrm{d}m_1 \mathrm{d}m_2}\frac{\mathrm{d}t}{\mathrm{d}z} \frac{\mathrm{d}E_s}{\mathrm{d}f_s},
\end{equation}
where $\rho_c$ is the critical density, and ${\mathrm{d}E_s}/{\mathrm{d}f_s}\propto (f |h(f)|)^2$ is the spectral energy density of a binary, computed at the source-frame frequency $f_s=f(1+z)$ ($f$ being the frequency at the detector). 
Our model's prediction for $\Omega_{\rm GW}$ should be compared with the $1\sigma$ power-law integrated curves~\citep{2013PhRvD..88l4032T} in \citet{LIGObackground} for the aLIGO/advanced Virgo network in the observing runs O1 (2015 to 16) and O5 (2020--2022), which represent the network's sensitivity to standard cross-correlation searches~\citep{PhysRevD.59.102001}  of  power-law  backgrounds. 

\section{Results}
\label{sec:results}
\begin{figure}
\centering
\includegraphics[width=0.47\textwidth,natwidth=8.4cm,natheight=4.6cm]{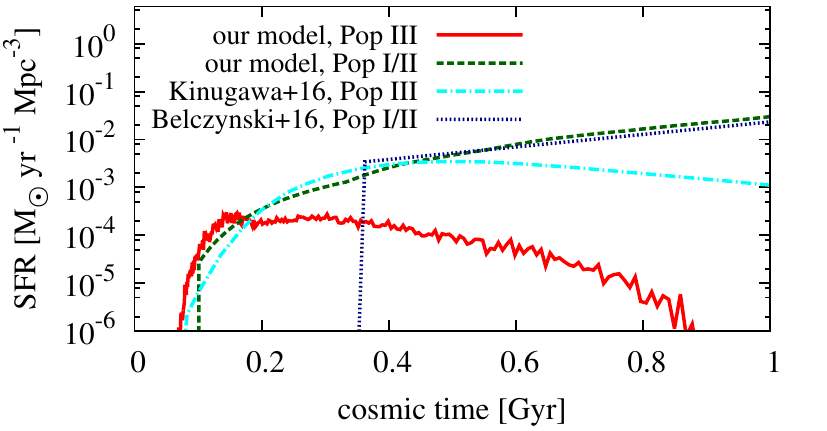}
\caption{Comparison of our SFR for the fiducial IMF with the models used by \citet{kinugawa16} (hereafter \citetalias{kinugawa16}) and \citetalias{belczynski16}. Our Pop I/II SFR, which is adopted from \citet{behroozi15}, is in good agreement with the corresponding SFR by \citetalias{belczynski16}. For the Pop~III stars, our self-consistent modelling yields a peak SFR that is about an order of magnitude lower than the value by \citetalias{kinugawa16}.}
\label{fig:SFR}
\end{figure}

In Fig. \ref{fig:SFR}, we compare our star formation rate (SFR) to other models. Our self-consistent Pop~III SFR, with a peak value of $\mathrm{SFR}_\mathrm{max}=2\times 10^{-4}\Msun \,\mathrm{yr}^{-1}\,\mathrm{Mpc^{-3}}$, is in compliance with \citet{visbal15}, who show that it cannot exceed a few $\times 10^{-4}$, without violating the constraints set by \citet{planck15}. \citetalias{kinugawa14} assume an SFR with a peak value of $\mathrm{SFR}_\mathrm{max}=3\times 10^{-3}\Msun \,\mathrm{yr}^{-1}\,\mathrm{Mpc^{-3}}$, which is about an order of magnitude higher than our result. The SFR is about the same for all our Pop~III IMFs, because we calibrate each model  to match $\tau$. The exact redshift evolution of the Pop~III SFR depends on the details of the treatment of reionization and metal enrichment  \citep[cf.][for a comparison]{johnson13}.

\begin{figure}
\centering
\includegraphics[width=0.47\textwidth,natwidth=8.4cm,natheight=8.4cm]{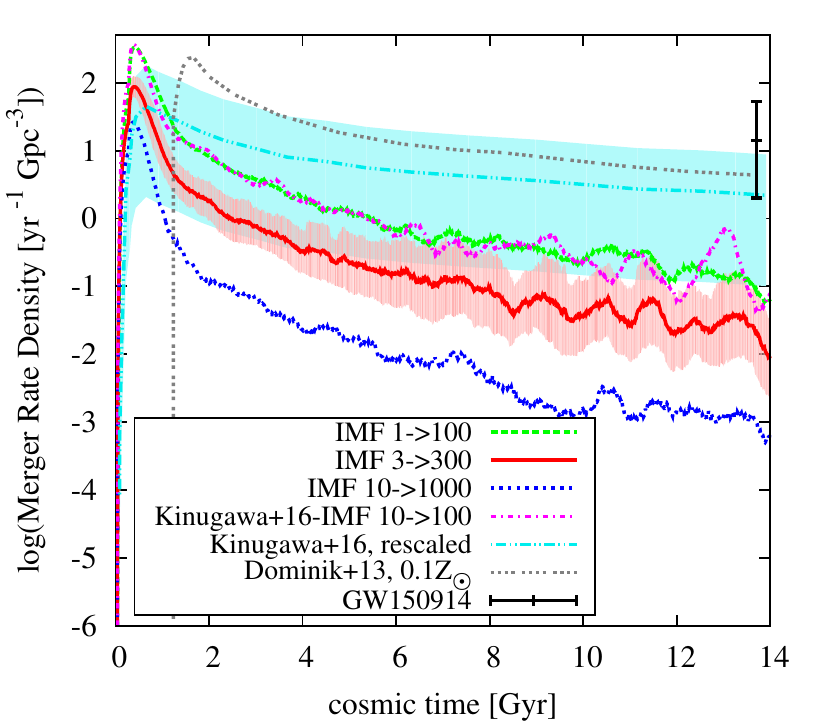}
\caption{Intrinsic merger rate densities (BH--BH, BH--NS, and NS--NS) for our models and comparison to the literature. For clarity, we show the statistical variance as shaded region only for our fiducial model. We plot the model by \citetalias{kinugawa16} for Pop~III remnants, (their fig.~22, model `under100, optimistic core merger'), rescaled to our $\mathrm{SFR}_\mathrm{max}$ with the corresponding systematic uncertainty. The model by \citet{dominik13} determines the merger rate density for stars at $10\%$ solar metallicity (their fig.~4). We also show the expected value at $z=0$ from the GW150914 detection \citep{LIGOrates}.}
\label{fig:rates}
\end{figure}

The intrinsic merger rate density of compact objects can be seen in Fig. \ref{fig:rates}.
To compare with \citetalias{kinugawa16} we rescale their SFR to our peak value, and run a case with their IMF (10--100).  Our values are about an order of magnitude lower than the rescaled prediction by \citetalias{kinugawa16} in the regime relevant for GW detections, i.e. mergers occurring at late cosmic times. This is to be ascribed to the different binary evolution models. Given that \citetalias{kinugawa14} study explicitly Pop~III star binary evolution, and we extrapolate models at higher metallicity, we conclude that our estimates are a lower limit to the GW detections of Pop~III binaries by up to an order of magnitude. We stress, however, that our Pop~III SFR is calculated self-consistently and reproduces the optical depth constraint set by \citet{planck15}, in contrast to \citetalias{kinugawa14,kinugawa16}.

Comparing models with different IMFs, the number of expected mergers is dominated by $M_\mathrm{max}$. This is because the remnants at low stellar masses are mostly NSs, which make only a small contribution to the overall merger rate. The merger rate density generally decreases with higher $M_\mathrm{max}$, because fewer stars (hence, binaries) form for a given total stellar mass. At face value, Pop~III stars do not yield a major contribution to the total merger rate density, but we recall that our estimates are likely lower limits. Crucially, due to their higher masses, Pop~III BH--BH mergers have strong GW signals, which boost their detection probability over lower mass BHs formed from later stellar generations.

Another essential question is whether we are able to discriminate mergers of primordial origin. The most massive remnant BHs for binaries at $0.1\,Z_\odot$ have a mass of $\sim\,42\Msun$ \citepalias{mink15}. All BHs with higher masses must be of primordial origin  \citep[though note that binary BHs with $M_{\rm tot}$ up to $\sim\,160 M_\odot$ may form in globular clusters;][]{,belczynski14,2016arXiv160202444R}. The minimal mass for Pop~III remnant BHs above the PISN gap is $M_\mathrm{PI} \approx 200\Msun$.
Since aLIGO will measure the  source-frame total mass for $M_{\rm tot}\gtrsim M_\mathrm{PI}$ within a 20\% uncertainty at the 2-$\sigma$ level (\citealt{graff15}, see also \citealt{veitch15,2016MNRAS.457.4499H}), we can use $M_\mathrm{PI}$ as the threshold for the unambiguous detection of a primordial BH.
For IMFs extending above $300\Msun$ about one BH--BH merger per decade can be unambiguously attributed to a Pop~III binary (Table \ref{tab:detection}). As discussed above, this is plausibly a lower limit because of the too efficient mass loss and hence, the underestimated remnant masses in our binary evolution model.

\begin{table}
 \centering
 \begin{tabular}{|c|c|c|c|c|}
  IMF &BH--BH&BH--NS&NS--NS&$m_1 > M_\mathrm{PI}$\\ 
  \hline 
  1--100& $5.3$ & $1.4 \times 10^{-2}$ & $7.2 \times 10^{-3}$ & $0$\\ 
  3--300& $0.48$ & $2.1 \times 10^{-3}$ & $8.1 \times 10^{-4}$ & $0.011$\\
  10--1000& $0.12$ & $2.4 \times 10^{-4}$ & $1.1 \times 10^{-5}$ & $0.089$\\ 
  \end{tabular} 
  \caption{Detection rates in events per year for aLIGO at final design sensitivity. Assuming a log normal distribution, we find a statistical scatter ($1 \sigma$) between different independent realizations of $(0.04-0.43)$\,dex, depending on the IMF and type of the merger. For the fiducial IMF (3--300), we expect about one Pop~III binary BH every two years and for the lower mass IMF (1--100), even up to 5 detections per year. The probability to detect a merger that can uniquely be identified as being of primordial origin ($m_1 > M_\mathrm{PI}$) is highest for the high-mass IMF (10--1000) with about one detection per decade. Inversely, the strong dependence of the detection rate on the IMF can be used to infer the upper mass limit for Pop~III stars.}
   \label{tab:detection}
\end{table}

\begin{figure}
\centering
\includegraphics[width=0.45\textwidth,natwidth=8.4cm,natheight=5.3cm]{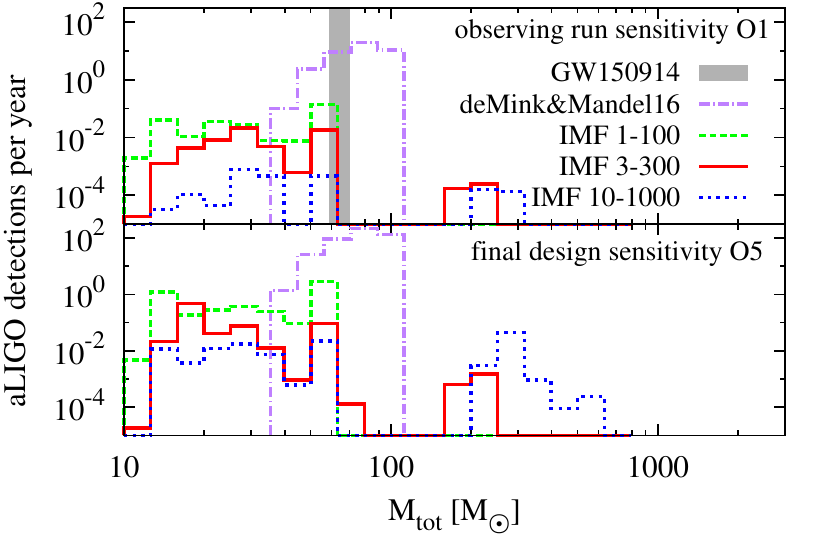}
\caption{Expected number of BH--BH merger detections per year as a function of the total binary mass for the current aLIGO sensitivity (top) and final design sensitivity (bottom). The mass range of GW150914 is indicated by the grey area.
With sufficient detections around $M_\mathrm{tot} \approx 300\Msun$, we could discriminate different Pop~III IMFs based on their GW fingerprint.}
\label{fig:histo}
\end{figure}

To distinguish the contributions to the detection rate by different generations of stars (also below $M_\mathrm{PI}$) we show the specific detection rates as a function of $M_\mathrm{tot}=m_1+m_2$ in Fig. \ref{fig:histo}.
We compare our three different IMFs with a model by \citet{mink16}. They determine the detection rate of Pop I/II stars for the chemically homogeneous evolutionary channel for binary black hole mergers, which dominates at $30\Msun \lesssim M_\mathrm{tot} \lesssim 100 \Msun$.
For a given $M_\mathrm{tot}$ the histogram enables to determine the probability that this event has a primordial origin. For GW150914, this probability is $\gtrsim1\%$ \citepalias[see][for different approaches]{belczynski16,woosley16}. For detections around $\sim\,300\Msun$, in addition to unambiguously establishing a Pop~III origin, one can even distinguish different Pop III IMFs by their number of detections over the next decades.

In our models, the highest stochastic GW background is produced by BH--BH mergers, and at $f\approx 25\,$ Hz (where the network is most sensitive), 
$\Omega_{\rm GW}(25\,\mbox{Hz})\approx 4\times 10^{-12}$--$\,1.4\times 10^{-10}$ (the range corresponds to different IMF choices; the inclusion or removal of resolved sources only makes a negligible difference). For comparison, the stochastic background at $25\,$Hz
inferred from GW150914 is $\Omega_{\rm GW}(25\mbox{Hz})=1.1^{+2.7}_{-0.9}\times 10^{-9}$ at 90\% confidence level, and the aLIGO/advanced Virgo network $1\sigma$ sensitivity (corresponding to SNR$\,=1$) for two years of observation at design sensitivity (O5) is $\Omega_{\rm GW}(25\mbox{Hz})=6.6\times 10^{-10}$. The stochastic background produced by Pop~III remnant mergers is therefore negligible at the relevant frequencies (compare \citealt{dvorkin16}, but see \citealt{inayoshi16}).

\section{Discussion}
We have estimated the GW fingerprint of Pop~III remnants on the aLIGO data stream. GWs have the potential to directly detect the remnants of the first stars, and possibly even to constrain the Pop~III IMF by observing BH--BH mergers with
 a total mass around $M_\mathrm{tot} \approx 300 \Msun$. The latter is the key to ascertain the impact of the first stars, due to their radiative and supernova feedback, on early cosmic evolution. The new GW window ideally complements other probes, such as high-$z$ searches for
energetic supernovae with the {\it James Webb Space Telescope (JWST)}, or 
stellar archaeological surveys of extremely metal-poor stars \citep{bromm13}.
We have developed a model which includes Pop~III star formation self-consistently, anchored, within the uncertainties, to the Planck optical depth to Thomson scattering. The main caveats in this study arise from the still uncertain Pop~III binary properties and the corresponding stellar binary evolution. 

We find a probability of $\gtrsim1\%$ that GW150914 originates from Pop~III stars, although this number may increase with improved future modelling. Crucially, the higher masses of the first stars boost their GW signal, and therefore their detection rate. Up to five detections per year with aLIGO at final design sensitivity originate from Pop~III BH--BH mergers. Approximately once per decade, we should detect a BH--BH merger that can unambiguously be identified as a Pop~III remnant. It is exciting that the imminent launch of the {\it JWST} nearly coincides with the first
direct detection of GWs, thus providing us with two powerful, complementary
windows into the early Universe.

GWs from BH binaries originating from Pop III stars may also be detectable by the { \it Einstein Telescope} \citep{sesana09} or (in their early inspiral) by eLISA \citep{2010ApJ...722.1197A,2016arXiv160206951S}, which would allow probing the physics of these systems with unprecedented accuracy.

\subsection*{Acknowledgements}

We thank P. Kroupa, S. Glover, M. Latif, D. Whalen, S. Babak, C. Belczynski, and the referee for helpful contributions. We acknowledge funding under the European Community's Seventh Framework Programme (FP7/2007-2013) via the European Research Council Grants `BLACK' under the project number 614199 (TH, MV) and `STARLIGHT: Formation of the First Stars' under the project number 339177 (RSK, MM), and  the Marie Curie Career Integration Grant GALFORMBHS PCIG11-GA-2012-321608, and  the H2020-MSCA-RISE-2015 Grant No. StronGrHEP-690904 (EB). RSK acknowledges support from the DFG via SFB 881, `The Milky Way System' (sub-projects B1, B2 and B8) and from SPP 1573 `Physics of the Interstellar Medium'. VB was supported by NSF grant AST-1413501.  AS gratefully acknowledges support through NSF grant AST-1211729 and by NASA grant NNX13AB84G. We thank the {\sc galform} team to make their code publicly available.

\bibliographystyle{mn2e}

\bibliography{GWpaper_sub3}

\bsp
\label{lastpage}

\end{document}